\documentclass[12pt,aps,floats]{revtex4}
\usepackage{amsmath,amsthm,amsfonts,graphicx,epsfig,mathrsfs}

\newcommand{\dbar}{\kern-.1em{\raise.8ex\hbox{ -}}\kern-.6em{d}}
 \font\tenrm=cmr10 
\begin{document}
\title{Topological swimming in a quantum sea}
\author{ J.E. Avron,  B. Gutkin and D.H. Oaknin\\
\tenrm\!  Department of
Physics, Technion, Haifa 32000, Israel\\
}
\begin{abstract}
We propose a quantum theory of swimming for swimmers that
are small relative to the coherence length of the medium. The
quantum swimming equation is derived from known results on quantum
pumps. For a one-dimensional Fermi gas at zero temperature we find
that swimming is topological: The distance covered in one swimming
stroke is quantized in half integer multiples of the Fermi wave length.
Moreover, one can swim without dissipation.
\end{abstract}
\maketitle

The theory of classical swimming  studies how a cyclic change in
the shape of a swimmer immersed in a fluid leads to a
change in its location. The theory is both elegant and practical
\cite{wilczek,childress} and has been applied to the swimming and
flying of organisms, robots \cite{purcel} and microbots
\cite{najafi, nature}.

A classical medium may be viewed as a limiting case of
quantum medium when the quantum coherence length is small compared
to the size of the swimmer and interference is negligible. Quantum
mechanics takes over once the swimmer is small enough. Our primary
goal here is to present a theory of swimming in quantum media.
We concentrate on swimming in a one dimensional
ideal Fermi gas at low temperatures but our methods are more
general and can be applied to higher dimensions. One dimension is
distinguished in that interference effects are especially strong
(and also leads to tractable and soluble models).

The swimmer, a q-swimmer, is an object with internal degrees of
freedom which, we assume, are slow and classical while the degrees
of freedom of the medium are fast and quantum. Swimming is
accomplished by the internal degrees of freedom, the controls,
undergoing periodic cycles.  An example of a q-swimmer might be a
molecule immersed in a quantum gas.  The control of the internal
configuration might be either external (through the application of
external fields) or internal (due to autonomous dynamics). 

Natural q-swimmers and artifical q-bots may be regarded as quantum
machines that propagate by pumping the quantum particles of the
ambient medium. This point of view reflects an intimate connection
between quantum pumps \cite{marcus} and q-swimming
which will play a role below.


Consider a simple model of a swimmer made of $n$ disconnected
spheres of radii $a_j$ immersed in either a classical or quantum
medium.  The swimmer can control the $n-1$ relative distances
$\ell_i=X_i-X_n, \ i=1,\dots,n-1$ between the centers $X_i$
of the spheres and the radii $a_j$.
Allow the swimmer to change adiabatically the control parameters
$a_j$ and $\ell_i$. The notion of
adiabaticity means that the velocities, e.g. $\dot X_j$, $\dot
a_j$, are small compared with the characteristic velocities of the
particles in the medium. By linear response, the force on the j-th
sphere is given by
\begin{equation}\label{lin-response}
f_j=-\sum_k \eta_{jk} \dot X_k-\sum_k \nu_{jk}\dot a_{k}+\sum_k F_{jk}.
\end{equation}
Here $F_{ji}=-F_{ij}$ are internal forces acting between $i$ and $j$ spheres and $\eta_{jk}$, $\nu_{jk}$ are coefficients, which, a-priori,
depend on the state of the swimmer (i.e. the relative distances
$\ell_{i}$ and the radii of the spheres $a_{j}$) and the nature of
the medium.


To derive an explicit form of the swimming equation one first needs to
make a choice how to designate the position of the swimmer, $X$.
For a swimmer made of $n$ disconnected pieces it is convenient to pick
$X = X_n$, the coordinate of one of the components.
The total force $\sum_{j}f_j$ acting on an adiabatic swimmer must vanish. (This
follows from the fact that the friction forces  are of first order in
the adiabaticity while the acceleration is second order.) This
constraint determines the swimming equation
\begin{equation}\label{swimming-general}
-\eta\,\dbar X=\sum_{i=1}^{n-1}\eta_i\,d\ell_i+\sum_{j=1}^{n}
\nu_j\,d a_{j}\,,
\end{equation}
where $\eta_k=\sum_{j=1}^n\eta_{jk}\,,\eta=\sum_{j=1}^n \eta_j\,,
\nu_k=\sum_{j=1}^n\nu_{jk}$. Swimming is manifestly geometric
being independent of the (time) parametrization of the swimming
stroke. The notation $\dbar X$ stresses that the position of the
swimmer will, in general, not integrate to a function on the space
of controls: $X$ will not return to its original values when the
controls undergo a cycle. 


Eq.~(\ref{swimming-general}) does not determine the
coefficients $\eta_i$, $\nu_j$  and $\eta$. In this sense, the
swimming equation, although general, is incomplete. We start by
deriving a new formula  for the quantum friction $\eta$ \cite{berry}.
$\eta$ expresses the friction acting on an idle q-swimmer while it
is being dragged at small velocity through the ambient quantum
medium. The formula is in the spirit of Landauer formula
\cite{landauer,imry}; It is expressed in terms of the scattering
data. As usual in the Landauer setting we assume a one dimensional
system. For the sake of simplicity, we focus on the two channel
case.

\begin{figure}[htb]
\hskip 0 cm
\includegraphics[width=4cm]{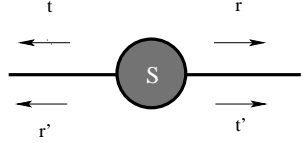}
\caption{\em A scatterer with two scattering channels:
$(r,t)$ and $(r',t')$ are the
reflection and transmission amplitudes}\label{pump}
\end{figure}

The most general on-shell scattering matrix of a one dimensional
time-reversal invariant scatterer anchored at the origin, can be
parameterized as follows \footnote{ Our convention differs from the usual
convention in mesoscopic physics by the interchange of rows.}:
\begin{equation}\label{scattering1d}
S_{0}\equiv \left(%
\begin{array}{cc}
  t & r' \\
  r&   t'  \\
\end{array}%
\right)=e^{i\gamma}\left(%
\begin{array}{cc} i \sin\theta  &   e^{-i\alpha}\cos\theta  \\
  e^{i\alpha}\cos\theta&  i \sin\theta\\
\end{array}%
\right).
\end{equation}
The reflection $r$, $r'$ and transmission $t$,  $t'$ amplitudes
are functions of three independent real parameters:
$\alpha,\gamma,\theta$.

The scattering matrix of a scatterer located at $X$ is related to
the scattering matrix located at the origin, $S_0$, by:
\begin{equation}\label{S}
S(X)=e^{iP X/\hbar}S_{0} e^{-iP X/\hbar},
\end{equation}
where  the on-shell  momentum matrix $P$ is defined by
\begin{equation}\label{momentum}
P=p(E)\left(%
\begin{array}{cc}
  1 & 0 \\
  0& - 1 \\
\end{array}%
\right),
\end{equation}
$p(E)=\sqrt{2mE}$ for electron gas and $p(E)=E/c$ for photons.

A dragged scatterer may be thought of as a pump. If the velocity
of dragging is small compared with the characteristic velocity of
the scattered particles, the theory of adiabatic pumps can be
applied. In particular, the rate of momentum transfer to the
ambient medium is
 \cite{ref:aegs}:
\begin{equation}\label{basic}
\dot{\cal P}(t)=-\frac 1 {2\pi\hbar} \int dE\,
\rho'(E)\,Tr_E\,\Big({\cal E}(E,t)\, P\,\Big),
\end{equation}
${\cal E}(E,t) = i\hbar \,\dot S\,S^*$ is called the energy shift matrix
\cite{martin}. It depends on a frozen on-shell scattering matrix and its time
derivative. Here and after $S^*$ denotes the Hermitian conjugate of $S$, $Tr_E$ denotes the trace on the scattering channels at
fixed energy $E$ and $\rho(E)$ gives the occupation of the
(scattering) states at energy $E$. If the ambient quantum gas is
at thermal equilibrium, $\rho(E)=( e^{\beta(E-\mu)}\pm 1)^{-1}$ is
the Fermi-Dirac or Bose-Einstein distribution.

In dragging a scatterer, the time dependence of $S$ comes solely
from the change of position, $X$. From Eq.~(\ref{S})
${\cal E}=\dot X [P,S]S^*\,$.
Eq.~(\ref{basic})  determines the force $f$ on the swimmer:
$f=\dot{\cal P}$.

Define, as usual, the friction coefficient, $\eta$, by $f=-\eta
\dot X$. Combined with Eq.~(\ref{basic}) we get
a Landauer type formula  for the quantum friction
\begin{eqnarray}\label{friction}
\eta=- \frac 1 {4\pi\hbar} \int dE\,
\rho'(E)\,Tr_E\,\Big( \,[P, S][P, S]^*\,\Big).
\end{eqnarray}
At thermal equilibrium $\rho(E)$ is a decreasing function of the
energy. This implies that $\eta$ is non-negative.

For a Fermi gas at  zero temperature $\rho'(E)=-\delta(E-E_F)$. In
the two channels case one then has
$\eta_F= \frac {2} {\pi \hbar} \, p^2(E_F)\, |r(E_F)|^2$.
The friction depends only on the momentum and reflection at the
Fermi energy, as one expects. Transparent objects are
frictionless.

To derive a q-swimming equation and fix the
coefficients of Eq.~(\ref{swimming-general}) we make use of the
elementary observation that swimming is dual to pumping. A turning
screw can be used to either pump or swim. The difference lies in
the setup: In a pump the external forces and torques adjust to
satisfy the constraints that the position and orientation of the
pump are fixed while in a swimmer the position and orientation of
the swimmer adjust to satisfy the constraint that there are no
external forces and torques. 

Assume that  no external forces are applied on a swimmer which can
control its scattering matrix. The rate of momentum transfer is
still given by Eq.~(\ref{basic}). Now, however, the energy shift
has two terms: ${\cal E}={\cal E}_{X}+{\cal E}_{0}$. The first
${\cal E}_{X}$ (given in Eq.~(\ref{basic})) arises from the
swimmer's change of location, while the second ${\cal
E}_{0}=i\hbar \,\dot S_{0}\,S_{0}^*$ comes from the swimming
stroke.

The total force acting on an adiabatic swimmer must vanish  (to
first order).
This means that $\dot {\cal P}$ vanishes and the equation of motion for
q-swimmers\footnote{The theory can be adapted to the
multidimensional case by imposing the constraint of vanishing
torque and momentum.}, is:
\begin{equation}\label{swimming}
\eta\dot X\ = \frac 1 {2\pi\hbar}\,\int dE \, \rho'(E)\,
Tr_E\,\Big({\cal E}_{0} P\Big)\,,
\end{equation}
where $\eta$ is the friction coefficient,  given in
Eq.~(\ref{friction}) and $P$ is given in
Eq.~(\ref{momentum}). In the two channels case the trace on the
right hand side is given by
\begin{equation}\label{rhs}
Tr_E\big({\cal E}_{0}P\big)= \hbar p(E) \,|r|^2\,{\rm Im}( \,d_t\log
(r/r'))\,.
\end{equation}


We shall now apply  the q-swimming equation, Eq.~(\ref{swimming}),
to swimming in a Fermi gas at zero temperature in one dimension.
This case is both simple and remarkable for, as we shall see,
q-swimming turns out to be topological. A small deformation of the
swimming stroke does not affect the swimming distance which is an
integer multiple of the Fermi wave length. This follows
immediately from
Eqs.~(\ref{swimming},\ref{rhs}) which
combine to give:
\begin{equation}\label{swimming-fermi-sea}
dX=\frac {\lambda_F} {8\pi}\, {\rm Im}\,\big(d\log (r/r')\big)
=\frac {\lambda_F} {4\pi}\,d  \alpha
\end{equation}
where $\lambda_F$ is the Fermi wave length. The fact that righthand  side is
an exact differential of the parameter $\alpha$ has two consequences: First,
to swim one must encircle the point where the scatterer is
transparent, $r=r'=0$, and second, the distance covered in a
stroke is quantized as a multiple of $\lambda_F/2$. The result is
general and does not depend on the specifics of the swimmers.



A swimmer will normally not have direct control on
parameters $\alpha$. Rather, it will  control some physical parameters that
will determine the scattering matrix (see examples below). Consider a swimmer with
two independent controls. A stroke is a closed path in the plane
of controls \footnote{Note that for $T=0$ the adiabatic limit breaks down at the  points $r=0$. Therefore, to make sense of adiabatic limit  a swimmer's
stroke must be chosen in a way it does not passes through transparency points.}. Since the reflection $r$ is a complex valued function
of the controls, one expects that by adjusting two controls, one
can find points where $r=0$ . With each such point of transparency
one can associate an (integer) index that counts how many times
$r$ rotates around the origin in one cycle around the point. We
call the index the {\em vorticity}. The swimming distance in a
closed stroke is proportional to the vorticity enclosed by the
path.  

Let us now consider two examples of topological swimmers.

\begin{figure}[htb]
\hskip 0.0 cm
\includegraphics[width=3.2cm]{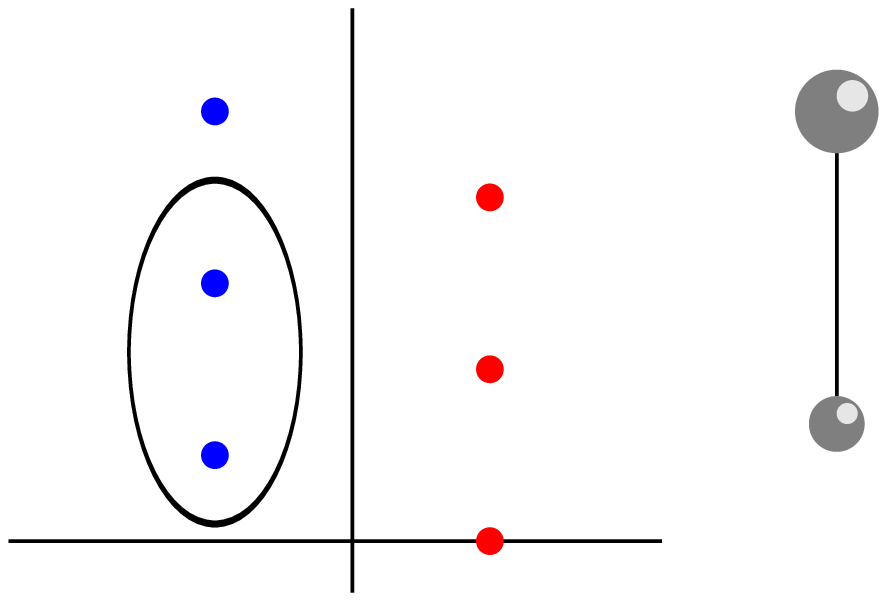}
\hskip 2.0 cm
\includegraphics[width=3.2cm]{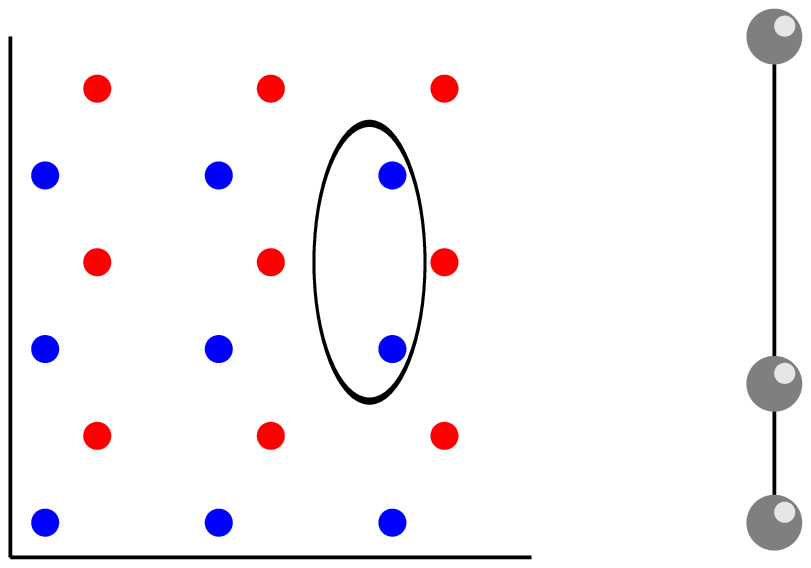}
\vskip -2.7 cm \hskip -6.0 cm  $\ell$ \vskip 2.0 cm \hskip -3.1 cm
$r_1/r_2$ \vskip -5.1 cm \hskip .5 cm
\vskip 2 cm \hskip 1.5 cm    $\ell_1$ \vskip 1.9 cm \hskip 6.0 cm
$\ell_2$ \vskip 0 cm
 \caption{\em The vortex structure, at $T=0$, of
Pushmepullyou (left) and the three linked spheres
(right). The swimming, in both cases, is topological: The distance
covered in one stroke is proportional to the sum of the
vorticities encircled by the path (the ellipses in the figures).
The red dots have vorticity $+1$ and the blue dots have vorticity
$-1$. }\label{fig:pushme}
\end{figure}

{\it Pushmepullyou:} Consider  `a molecule' made of two scatterers.
The scattering matrices associated with each scatterer have
$\alpha_j=\gamma_j=0$ for both scatterers and $r_1=\cos\theta_1$,
$r_2=\cos\theta_2$.
The two scatterers are separated by distance $\ell$, see
Fig.~(\ref{fig:pushme}). The swimmer can control $\ell$, and the
ratio $r_1/r_2$. The total scattering matrix of the swimmer can
then be computed by considering the multiple scattering processes
between the two scatterers. A computation yields for the zeros of
total reflection of $r$  the solutions of:
\begin{equation}\label{pair_ladders}
e^{2ik\ell} r_1-r_2=0.
\end{equation}
It follows that the vortices occur when $r_1=\pm r_2$ and
the distance $2\ell$ is an integer (half and integer) multiple of
wavelengths. Since the zeros are simple the vorticities are $\pm 1$.

{\it Three linked spheres}: Consider a swimmer (proposed by
Najafi and Golestanian in \cite{najafi} in a different context),
made of three identical scatterers on a line, separated by
distances $\ell_1$ and $\ell_2$, see Fig.~\ref{fig:pushme}. Take
$\alpha_j=\gamma_j=0$ and $r_j=\cos\theta$, independent of
$j=1,2,3$. The swimmer, a vibrating ``trimer'', now controls only
the two distances $\ell_1, \ell_2$. A computation shows that the
total reflection $r$ vanishes provided
\begin{equation}\label{pair}
e^{i2k\ell_1}+
e^{-i2k\ell_2}-\cos^2\!\theta \, e^{i2k(\ell_1-\ell_2)}-1=0.
\end{equation}
The vortices evidently occur  on a pair of lattices in  the plane
$(\ell_1,\ell_2)$, see Fig.~\ref{fig:pushme}. Once again, since
the zeros are simple, the vorticities are $\pm 1$.

\begin{figure}[htb]
\hskip 0 cm
\includegraphics[width=5.0cm]{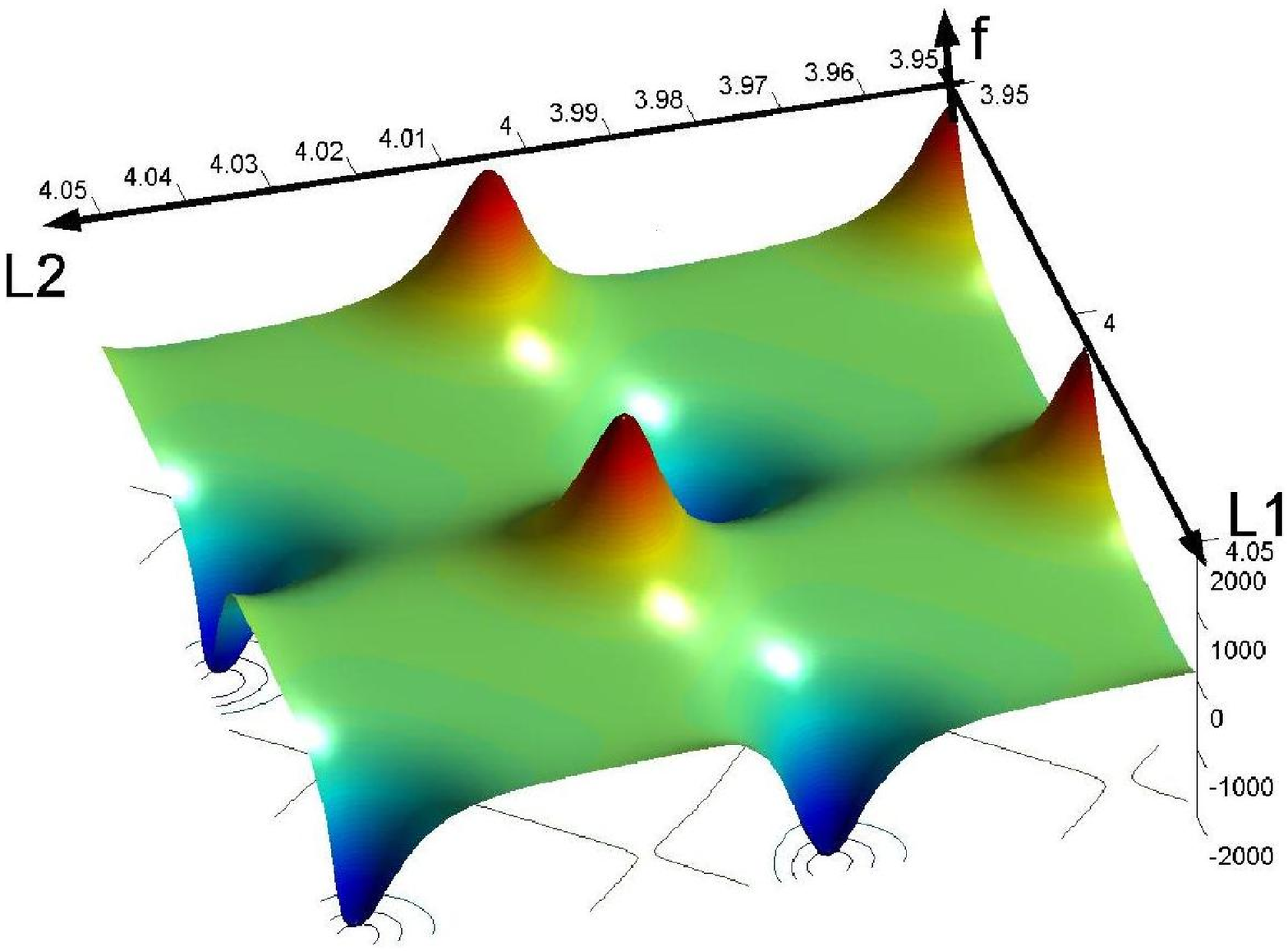}
\includegraphics[width=5.0cm]{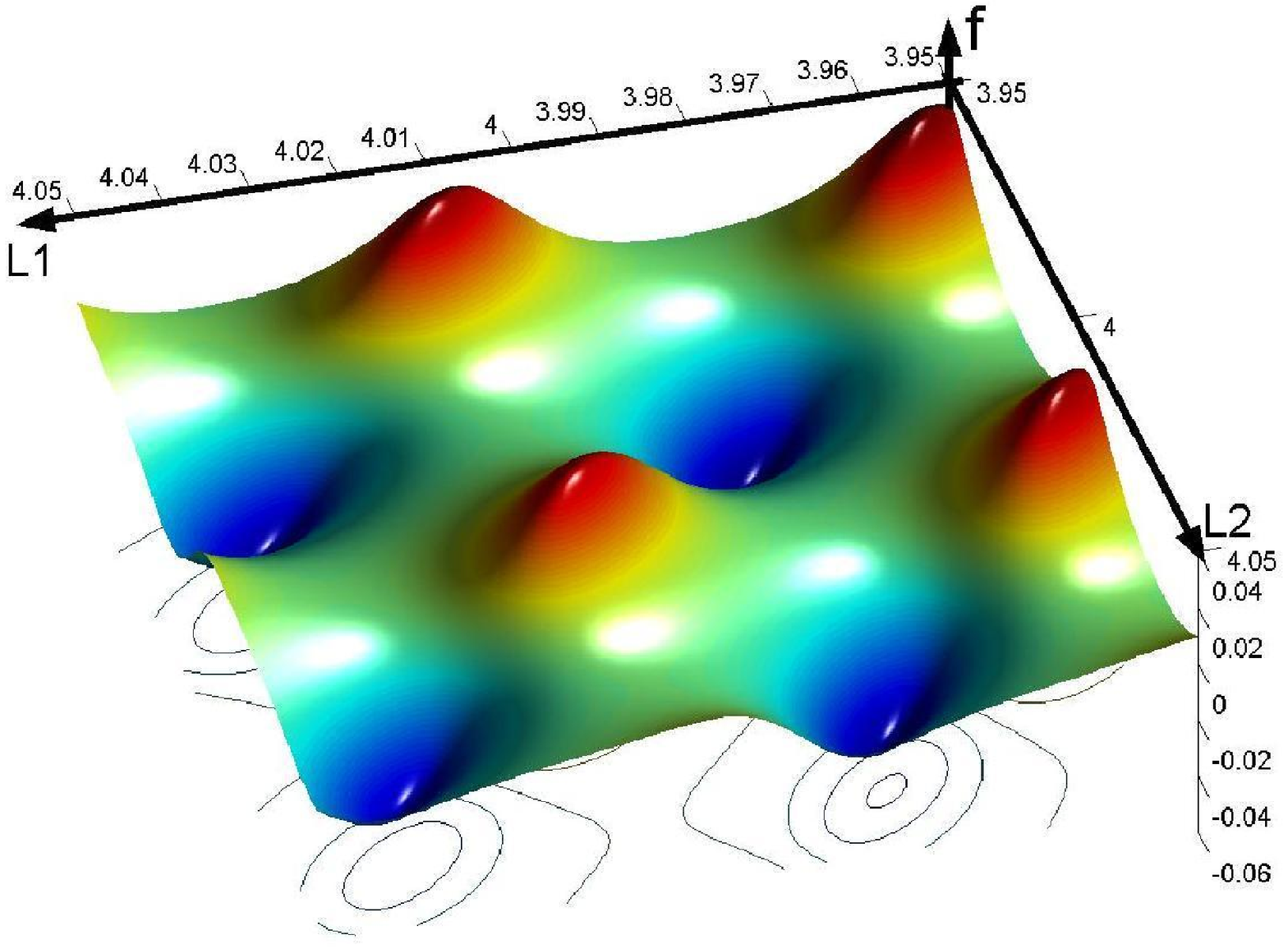}
\caption{\em  The curvature 
 for the three linked spheres, shown in a 3D plot as a
function of control parameters at low temperature,
$T=T_0/4$, (above) and high temperature, $T=4T_0$, (below). The
height of the peaks at the lower temperature is about a million
times the height of the peaks at the higher temperature. The curvature 
vanishes as the temperature gets higher. The distance 
covered in one stroke is the integral of the curvature over the surface 
in the space of controls enclosed by the stroke. 
}\label{fig:finite_temp}
\end{figure}
\begin{figure}[htb]
\hskip 0 cm
\includegraphics[width=5.0cm]{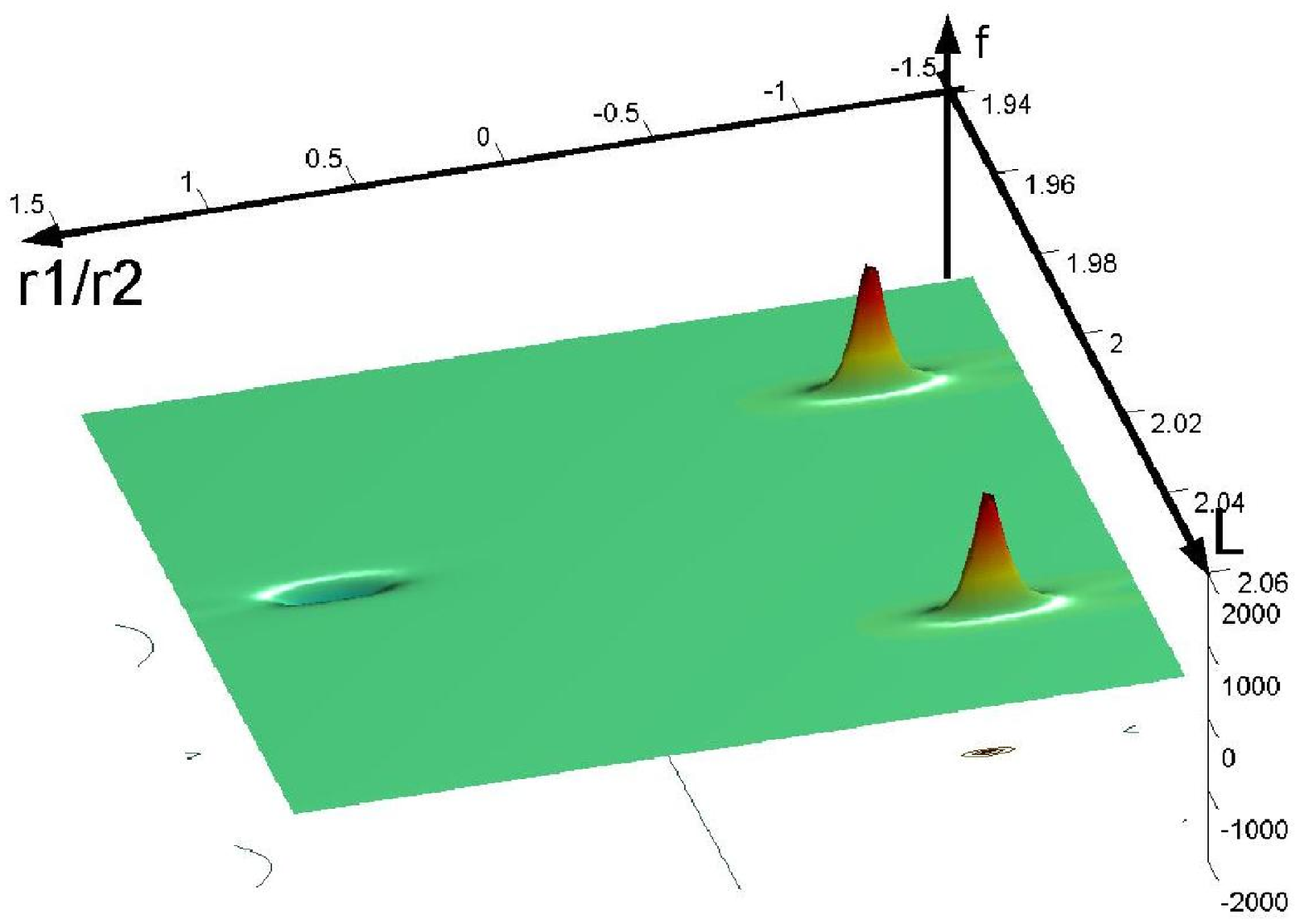}
\includegraphics[width=5.0cm]{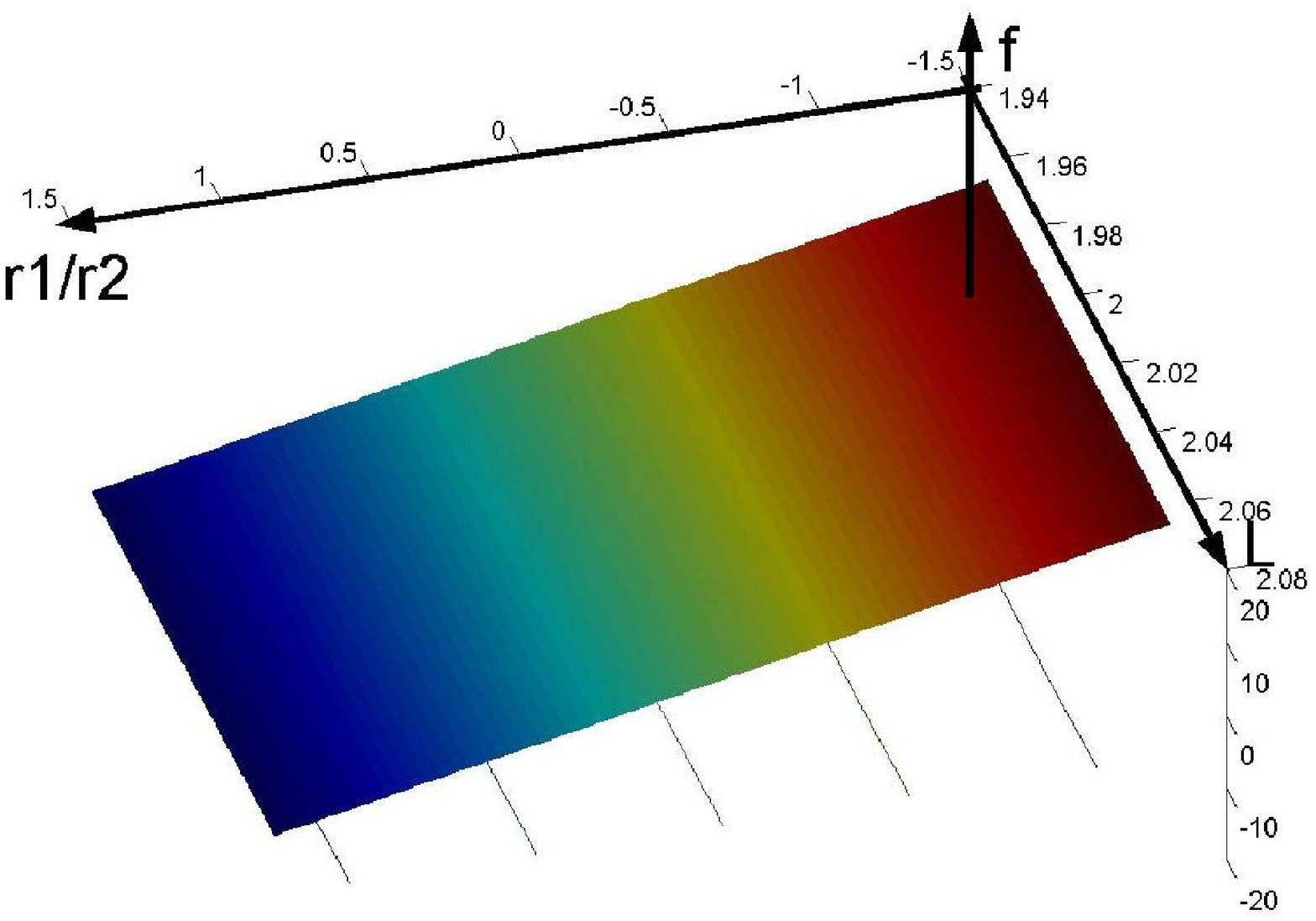}
\caption{\em  The curvature at finite
temperatures for Pushmepullyou, shown in a  3D plot as a function
of control parameters.  The upper picture corresponds to the case of low 
temperature. The case of high temperature is shown in the lower picture. 
In contrasts with the three linked spheres, for Pushmepullyou the curvature 
does not vanish at high temperatures. }\label{fig:finite_temp_pmpy}
\end{figure}

In geometrical language,  Eqs.~(\ref{swimming},\ref{swimming-fermi-sea})  define a
connection $\dbar X$ in the space of controls.
The distance covered by the swimmer in a one stroke $\mathcal{C}$
is then given by $\Delta X=\int_{\mathcal{C}}\dbar X$. When there
are only two control parameters $(x,y)$ (as in the examples
above), an application of the Stokes formula gives for the
displacement $\Delta X=\int\!\!\!\int f(x,y)\,dx\,dy $, where
$f(x,y)$ is the (scalar) {\em curvature}, and the domain of
integration has the boundary ${\mathcal{C}}$. A plot of the
curvature $f$ is often an efficient way to describe a swimmer, see
e.g., Figs.~\ref{fig:finite_temp},\ref{fig:finite_temp_pmpy}.

It is instructive
to see how topological Fermi swimmers are affected by temperature.
At temperature $T$ a region proportional to $T$ near the Fermi
energy will contribute to the integral in Eq.~(\ref{swimming}).
This makes the curvature a smooth function (rather than a
collection of delta functions) on the space of controls. The total
curvature enclosed in a path will now depend smoothly on the path.
The temperature scale is determined by $T_0= h^2/4mk_B
\lambda_F\ell $ where $k_B$ is Boltzmann constant and $m$ the mass
of the scattered particle. At this temperature the support of $f$
near a vortex  becomes comparable with the distance between
vortices. From the definition of $T_0$ it follows that the larger
the swimmer (the larger $\ell$) the more sensitive it is to
temperature.

Pushmepullyou and the three linked spheres have an essentially
different behavior at high temperatures. Since the neighboring
vortexes of three linked spheres (at $T=0$) are of opposite sign,
the smearing at high temperatures leads to vanishing curvature at
high temperatures, see Fig.~\ref{fig:finite_temp}. Thus the three
linked spheres do not swim effectively  at high temperatures. On
the other hand, the smearing of the the vortices of Pushmepullyou
(at $T=0$) does not lead to  mutual cancellation,  see
Fig.~\ref{fig:finite_temp_pmpy}. That means Pushmepullyou can swim
effectively also at high temperatures.


In the course of its motion a swimmer will, in general, transfer
energy to the medium and dissipate energy. As we shall now show,
swimming without dissipation is possible in one dimensional Fermi
gas at zero temperature.

We define the dissipation, $\dot D$, as the difference between the
outgoing and incoming energy current. From
\cite{ref:aegs}:
\begin{equation}\label{dissipation1}
\dot D=\frac{\hbar}{4\pi}Tr_{E_F}\big(\dot S \,\dot
S^*\big)\geq 0.
\end{equation}
One of the consequences of this relation is that a non-dissipating
swimmer is indistinguishable from a static scatterer (since $\dot
S=0$). Clearly, there is no dissipation in the former, and there
should therefore be no dissipation in the latter.

To see why it is possible to swim without dissipation, write
Eq.~(\ref{dissipation1}), (in the two channel case), in the form:
 \begin{equation}\label{dissipation2}
\dot D=\frac{\hbar}{2\pi}\big((\dot\alpha - 2 k_F  \dot
X)^2\,\cos^2\theta+ \dot \theta^2+\dot\gamma^2 \big).
\end{equation}
The first term vanishes by Eq.~(\ref{swimming-fermi-sea}). Thus if
$\dot\theta=\dot\gamma=0$ there is no dissipation. We call a
swimming without dissipation ``super-swimming''.

A super-swimmer needs a larger space of controls to ensure that
$\dot \theta =\dot\gamma=0$. (For instance, to make a
super-swimmer out of Pushmepullyou one needs to control also the
overall phase $e^{i\gamma}$ of the two individual scatterers.)
Super-swimmers are, in general, not transparent and one still
needs to invest power to drag them. Only when the super-swimmer
swims on its own, there is no transfer of energy to the ambient
medium.

We note that no-dissipation, $\dot S=0$, implies
Eq.~(\ref{swimming-fermi-sea}) and hence implies quantization.

{\bf Acknowledgment} This work is supported in parts by the EU
grant HPRN-CT-2002-00277, the ISF, and the fund for the
promotions of research at the Technion.



\begin{thebibliography}{10}

\bibitem{wilczek}  E.M. Purcell, Life at low Reynolds numbers, Am.
J. Physics 45, 3-11 (1977); A. Shapere and F. Wilczek,
J.\ Fluid Mech., {\bf
198}, 557-585 (1989) 
\bibitem{childress} S.~Childress, Mechanics of Swimming and Flying,
(Cambridge University Press, Cambride, 1981);
J. R. Blake, Math. Meth. Appl. Sci. 24, 1469 (2001).
\bibitem{purcel}L.E. Becker, S.A. Koehler, and H.A. Stone, J.
Fluid Mech. 490 , 15 (2003); E.M. Purcell, Proc. Natl. Acad. Sci.
94 , 11307-11311 (1977). 
\bibitem{najafi} A. Najafi and R. Golestanian, Phys. Rev. {\bf E69}
(2004) 062901, cond-mat/0402070 
\bibitem{nature} R. Dreyfus, J. Baudry, M.L. Roper, M. Fermigier, 
H.A. Stone, J. Bibette, Nature 
{\bf 437}, 862 - 865 (2005).
\bibitem{marcus} M. Switkes, C.M. Marcus, K. Campman, A.G. Gossard,
Science, 
{\bf 283}, 1907 (1999);P. W. Brouwer, Phys.\ Rev.\ B {\bf
63},121303 (2001);
P.W. Brouwer,
Phys.\ Rev.\ B {\bf 58}, 10135 (1998). 

\bibitem{berry} M.V. Berry and J.M. Robbins, Proc.~Roy.~Soc. {\bf
422} 659 (1993); D.~Cohen, Phys.~Rev.~B {\bf 68}, 155303 (2003)

\bibitem{landauer} R. Landauer, IBM J. Res. {\bf 1}, 233 (1957);
M. B\"uttiker, Phys.\ Rev.\ Lett.\ {\bf 57}, 1761 (1986)
\bibitem{imry} Y. Imry, {\em Introduction to mesoscopic physics}, Oxford University
Press (1997).



\bibitem{martin} P.A. Martin, M. Sassoli de Bianchi, J.\ Phys.\ A 28, 2403 (1995).

\bibitem{ref:aegs} J. Avron, A. Elgart, G.M. Graf and L. Sadun, J.\ Stat.\
 Phys. {\bf 116}, 452-473 (2004)

\bibitem{ref:wisdom} J. Wisdom,
Science {\bf  299},
1865-1869, (2003)
\bibitem{thouless} D.~J.~Thouless, Phys.\ Rev.\ B {\bf 27}, 6083 (1983);
Q. Niu, Phys.\ Rev.\ Lett.\ {\bf 64}, 1812 (1990).




\end{thebibliography}
\end{document}